\documentclass[lettersize,journal]{IEEEtran}

\usepackage{cite}
\usepackage{amsmath,amssymb,amsfonts}
\usepackage{algorithmic}
\usepackage{graphicx}
\usepackage{multirow}
\usepackage{url}
\usepackage{tabularx}
\usepackage{array}
\usepackage{float}

\newcolumntype{C}[1]{>{\centering\arraybackslash}m{#1}}
\floatstyle{ruled}
\newfloat{algorithm}{tbp}{loa}
\floatname{algorithm}{Algorithm}

\def\BibTeX{{\rm B\kern-.05em{\sc i\kern-.025em b}\kern-.08em
 T\kern-.1667em\lower.7ex\hbox{E}\kern-.125emX}}

\begin{document}

\title{An LLM-Assisted AutoML Framework for Intrusion Detection in IoT Networks}

\author{Li Yang,~\IEEEmembership{Member,~IEEE}%
\thanks{Li Yang is with the Faculty of Business and Information Technology, Ontario Tech University, Oshawa, ON L1G 0C5, Canada, and also with the Department of Electrical and Computer Engineering, Western University, London, ON N6A 3K7, Canada (e-mails: li.yang@ontariotechu.ca; lyang339@uwo.ca).}}

\maketitle

\begin{abstract}
Internet of Things (IoT) systems are increasingly deployed in smart homes, transportation, energy systems, and critical infrastructure. This broad connectivity improves service intelligence, but also enlarges the attack surface of IoT networks. Machine Learning (ML)-based Intrusion Detection Systems (IDSs) are widely used to identify malicious network threats and protect IoT systems, but developing effective ML-based IDS models often requires human expertise and repeated manual decisions on many procedures, including data pre-processing, feature selection, model selection, and hyperparameter tuning. Automated Machine Learning (AutoML) reduces this burden by automating steps of the ML pipeline using optimization techniques, but conventional AutoML methods can consume substantial optimization time because they explore broad candidate model families and large hyperparameter spaces. This paper proposes a Large Language Model (LLM)-assisted AutoML framework for IoT intrusion detection. The proposed framework uses an LLM as a policy generator that converts dataset profiles into bounded and validated AutoML policies for automated data balancing, automated feature engineering, and Combined Algorithm Selection and Hyperparameter Optimization (CASH). Under an equal 10-trial budget, the proposed LLM-assisted policy achieves higher weighted test F1-score than traditional AutoML using the Tree-structured Parzen Estimator (TPE) on both datasets, reaching 99.680\% on CICIDS2017 and 99.186\% on IoTID20. Relative to the broader 30-trial Traditional AutoML-TPE baseline, the 10-trial proposed method reduces optimizer time by 63.7\% and 49.9\%, respectively, while achieving slightly higher F1-score. These results show that a bounded LLM policy can improve the quality of a low-budget AutoML search while retaining a clear efficiency advantage relative to a larger conventional search budget.
\end{abstract}


\section{Introduction}
\IEEEPARstart{T}{he} Internet of Things (IoT) refers to the integration of sensing, computing, and communication capabilities into physical devices that can collect and exchange data over networked environments \cite{iot1,ieeeaccess_iot_review}. IoT has enabled many important applications, such as smart homes, intelligent transportation, smart grids, and critical infrastructure management \cite{automl_iot}. However, the increased connectivity that supports IoT applications also increases cybersecurity risks. Resource-constrained IoT devices often have limited processing power and security mechanisms, making them vulnerable to various cyber-attacks, such as Denial of Service (DoS), botnets, scanning, and Man-in-the-Middle (MITM) attacks \cite{iot1,bagaa_iot_security}. For example, in Electric Vehicle Charging Systems, connected chargers, backend servers, and smart grid interfaces improve charging intelligence, but they also expose charging operations and user information to cyber-attacks \cite{iot_evcs_survey,tinyml_evcs}.

Intrusion Detection Systems (IDSs) are essential security mechanisms for IoT systems because they analyze network traffic and detect malicious attacks that bypass preventive controls such as authentication and firewalls \cite{ids1,muthanna_iot_ids}. Machine Learning (ML)-based IDSs have become a dominant approach because ML models can learn complex relationships between network-traffic features and attack classes and generalize these patterns to unseen samples \cite{iot_evcs_survey,fatani_iot_ids}. However, developing an effective IDS remains challenging. IDS datasets are usually high-dimensional, multi-class, and highly imbalanced, and model performance usually depends strongly on the ML development process, including data pre-processing, feature selection, model selection, and hyperparameter tuning \cite{ids1}. Traditional ML development is constrained by human expertise and design bias, which can lead to suboptimal performance and insufficient generalizability. 

Automated Machine Learning (AutoML) has emerged as a promising approach to address these development challenges because it can automate key stages of the ML pipeline, including Automated Data Pre-processing (AutoDP), Automated Feature Engineering (AutoFE), automated model selection, and Hyperparameter Optimization (HPO) \cite{ids1}. This makes AutoML attractive for autonomous cybersecurity, especially in future networks where zero-touch networking principles aim to reduce human intervention in monitoring, optimization, and protection \cite{ids1}. The joint problem of automated model selection and HPO is commonly formulated as Combined Algorithm Selection and Hyperparameter Optimization (CASH), a central problem in AutoML \cite{automl_iot,hpo_theory}. Nevertheless, conventional AutoML also has limitations. The AutoML search process can be computationally expensive due to broad search spaces, but defining an effective search space remains a difficult problem and often requires domain knowledge \cite{ids1}. A generic AutoML process may spend many trials evaluating candidate models or hyperparameter regions that are unlikely to provide a favorable trade-off for a specific network security dataset. This challenge is especially important for IDS development in IoT systems, where efficient and resource-aware model construction is often needed \cite{automl_iot}.

Recent work has begun to use Large Language Models (LLMs) at specific stages of ML automation. Context-Aware Automated Feature Engineering (CAAFE) uses an LLM to generate context-aware feature transformations for tabular data \cite{caafe}, while Large Language Models to Enhance Bayesian Optimization (LLAMBO) integrates LLM guidance into Bayesian optimization and hyperparameter search \cite{llambo}. Broader studies have also examined LLMs as assistants for ML workflow design and AutoML \cite{llm_workflow_survey,llm_automl_risks}. These results motivate the use of LLMs for search-space design, but unrestricted agents may also produce invalid code, unsupported operations, or inconsistent configurations \cite{llm_automl_synergy}. For IDS development, we therefore restrict the LLM to a fixed AutoML policy schema. It proposes bounded choices from a compact dataset profile, while validated ML components perform data processing, model training, and evaluation. The central question in this work is whether such a bounded policy can outperform a conventional predefined AutoML policy under the same low-budget setting, and whether the resulting compact search can remain competitive with a broader high-budget search.

Therefore, this paper proposes an LLM-assisted AutoML framework for IoT intrusion detection. A local Llama 3.1 8-billion-parameter (8B) model \cite{llama3} receives an aggregate dataset profile and returns a structured policy for AutoDP-based balancing, AutoFE, and CASH optimization. The policy is checked against predefined operators, parameter bounds, and fallback rules before it is executed. Model selection remains empirical: the validated pipeline is evaluated by stratified cross-validation on the training partition, and the hold-out set is used only for the final assessment. Local execution keeps the profiling information within the experimental environment and allows the same LLM version and decoding settings to be reused across experiments. In this design, the LLM shapes a compact AutoML policy but does not replace the deterministic training and validation procedures.

Prior studies have examined LLM-assisted feature engineering and hyperparameter optimization, whereas this work focuses on a validated local policy interface that jointly coordinates AutoDP, AutoFE, and CASH for IoT IDS development. To the best of our knowledge, no prior IoT IDS framework has combined these three AutoML stages through a bounded LLM policy while leaving model fitting and scoring to deterministic and validation-driven procedures. The main contributions of this paper are summarized as follows:
\begin{enumerate}
\item An LLM-assisted AutoML framework\footnote{The complete code will be available upon paper acceptance at: \url{https://github.com/LiYangHart/LLM-Assisted-AutoML-For-Intrusion-Detection}} is proposed for IoT intrusion detection, in which a local LLM generates bounded policies for AutoDP, AutoFE, and CASH optimization. The framework is designed to improve the quality of a low-budget AutoML search while reducing unnecessary exploration of a broad search space.
\item A validated policy-execution mechanism is developed to convert LLM-generated suggestions into executable AutoML configurations, preventing unsupported model families and invalid ranges.
\item A detailed AutoML pipeline is designed using partial adaptive oversampling, multi-stage feature selection with retention safeguards, and CASH based on Bayesian Optimization (BO) with the Tree-structured Parzen Estimator (TPE), denoted BO-TPE, using a compact gradient-boosted-tree search policy for the proposed branch.
\item The proposed framework is evaluated on public benchmark cybersecurity datasets, CICIDS2017 \cite{cicids2017} and IoTID20 \cite{iotid20}, and compared with traditional ML baselines, state-of-the-art AutoML and optimized ML models, and ablated LLM-assisted variants.
\end{enumerate}

The remainder of this paper is organized as follows. Section \ref {S2} reviews optimized and automated ML-based IDSs and recent LLM-assisted AutoML research. Section \ref {S3} presents the proposed LLM-assisted AutoML framework in detail. Section \ref {S4} describes the experimental setup and discusses the results. Section \ref {S5} concludes the paper and outlines future research directions.

\section{Related Work} \label{S2}
\subsection{Optimized and Automated ML-based IDSs} 
ML-based IDSs have been widely used for network and IoT security \cite{ids1,mth_ids}. Supervised learning methods distinguish benign and malicious traffic by learning decision boundaries from labeled network-traffic records \cite{ids1}. Tree-based models are frequently adopted because network traffic datasets are usually tabular and nonlinear, while recent evidence also shows that tree-based models often remain highly competitive with deep learning on typical tabular data \cite{tabular_tree}. Jin \textit{et al.} \cite{lightgbm_ids} proposed a real-time IDS that uses the Light Gradient Boosting Machine (LightGBM) with a parallel intrusion detection mechanism to improve detection efficiency on large-scale traffic data. Fernando \textit{et al.} \cite{iot_idsai} introduced a new IoT intrusion-detection dataset and examined ML model generalization for IoT communication intrusion detection. Deep learning methods can also achieve strong detection performance, although tree-based models often remain competitive on tabular datasets and generally require less architectural design \cite{tabular_tree,transfer_cnn_iov}.

Optimized ML-based IDSs improve conventional ML by tuning models, selecting features, or constructing ensembles. Elmasry \textit{et al.} \cite{pso_lstm_ids} proposed a double Particle Swarm Optimization (PSO) method to evolve the Long Short-Term Memory (LSTM) model for network intrusion detection. Naeem \textit{et al.} \cite{ga_cnn_ids} developed a multi-class vehicle-security framework that combines deep transfer learning-based Convolutional Neural Network (CNN) with Genetic Algorithm (GA)-based optimization.

AutoML-based IDSs further reduce manual effort by automating the ML model optimization procedures.
Khan \textit{et al.} \cite{oe_ids} proposed the Optimized Ensemble Intrusion Detection System (OE-IDS), an AutoML-based soft-voting ensemble model for network intrusion detection. Singh \textit{et al.} \cite{automl_id} proposed Automated Machine Learning for Intrusion Detection (AutoML-ID), which uses AutoML to select and tune IDS models for wireless sensor networks. Yang and Shami \cite{automl_ccs} proposed an AutoML-based autonomous IDS framework that integrates automated data pre-processing, feature engineering, model selection, and hyperparameter optimization for intrusion detection.

Although optimized ML-based and traditional AutoML-based IDSs have improved detection performance, several limitations remain. Many optimized IDSs still rely on manually defined search spaces, while general AutoML methods may explore broad candidate spaces and consume considerable computational time. Moreover, most existing AutoML-based IDSs mainly rely on empirical search and provide limited support for using dataset characteristics to guide the AutoML process before optimization. These limitations motivate an LLM-assisted AutoML design in which a compact dataset profile is used to propose a targeted and validated search policy for IDS development.

\newcolumntype{L}{>{\raggedright\arraybackslash}X}

\begin{table}[tb]
\centering
\renewcommand{\arraystretch}{1.25}
\setlength{\extrarowheight}{1pt}
\setlength{\tabcolsep}{2.8pt}
\caption{Comparison of related IDS and AutoML research categories.}
\footnotesize
\begin{tabularx}{0.99\columnwidth}{|>{\centering\arraybackslash}p{1.45cm}|>{\centering\arraybackslash}p{1.20cm}|L|L|}
\hline
\textbf{Category} 
& \textbf{Related Works} 
& \textbf{Main Contribution} 
& \textbf{Main Limitation / Novelty} \\ \hline

Traditional ML-based IDSs 
& \cite{lightgbm_ids,iot_idsai} 
& Detect attacks from network traffic using manually designed ML models. 
& Require manual pre-processing, model selection, and tuning; provide the IDS foundation. \\ \hline

Optimized ML-based IDSs 
& \cite{pso_lstm_ids,ga_cnn_ids} 
& Improve IDS performance through feature selection, ensembles, or HPO. 
& Search design and objectives are usually manually defined; motivate optimized IDS development. \\ \hline

Traditional AutoML-based IDSs 
& \cite{automl_id,automl_ccs} 
& Automate model selection, HPO, and other ML pipeline stages. 
& Broad search spaces may increase runtime; provide the AutoML backbone. \\ \hline

LLM-assisted ML/AutoML 
& \cite{caafe,llambo} 
& Use LLMs to suggest ML workflow and optimization choices. 
& Free-form LLM outputs require validation; motivate bounded policy generation. \\ \hline

Proposed framework 
& This paper 
& Uses local LLM policy generation for AutoDP, AutoFE, and CASH in IDS. 
& Integrates LLM guidance with validated AutoML execution for efficient and reproducible IDS development. \\ \hline

\end{tabularx}
\label{liter}
\end{table}

\subsection{LLM-Assisted ML Workflow Design}
LLMs have recently been studied for data analysis, ML workflow construction, and AutoML search \cite{llm_workflow_survey,llm_automl_risks}. Two direct examples are CAAFE, which uses an LLM to iteratively generate context-aware feature-engineering code for tabular data \cite{caafe}, and LLAMBO, which introduces LLM-based components for warm-starting, surrogate modeling, and candidate sampling in Bayesian optimization \cite{llambo}. These methods show that LLMs can contribute useful prior information to individual stages of an AutoML process. However, free-form feature generation or direct LLM control of an optimization loop is difficult to validate in a security experiment, where data partitioning, pre-processing, and model scoring must remain reproducible.

LLMs have also been integrated more directly into intrusion-detection and security-analysis workflows. IDS-Agent uses an LLM-driven agent with specialized IDS tools, memory, and external knowledge to reason over network traffic and produce explainable intrusion decisions \cite{ids_agent}. HuntGPT combines a Random Forest anomaly detector with explainable-AI components and an LLM-based conversational interface to present detected threats in an analyst-readable form \cite{huntgpt}. These systems place the LLM in the detection or interpretation loop. In contrast, the proposed framework keeps the deployed IDS conventional: the LLM is used only during offline AutoML planning, does not inspect hold-out or live traffic, and does not produce final intrusion labels. Instead, it operates at the policy level, jointly guiding data balancing, feature selection, and CASH within a predefined and validated decision space, while deterministic ML procedures perform model fitting, validation, and final prediction.

This design allows several interconnected AutoML stages to be adapted together rather than optimizing an isolated component, while also reducing unnecessary search and improving the efficiency of model development. The use of a local LLM further limits external data exposure and avoids dependence on online services. Overall, the proposed framework combines flexible LLM-assisted policy generation with a controlled and practical AutoML pipeline for IoT intrusion detection. Table~\ref{liter} places this design relative to existing IDS and AutoML research.

\section{Proposed LLM-Assisted AutoML Framework} \label{S3}
\subsection{System Overview}

The proposed framework uses LLM-guided AutoML planning to construct an optimized IDS with a smaller search cost. Rather than allowing the LLM to generate executable programs, its output is restricted to a bounded AutoML policy. All data processing, feature selection, model fitting, and scoring are performed by predefined operators.

\begin{figure*}
 \centering
 \includegraphics[width=17.5cm]{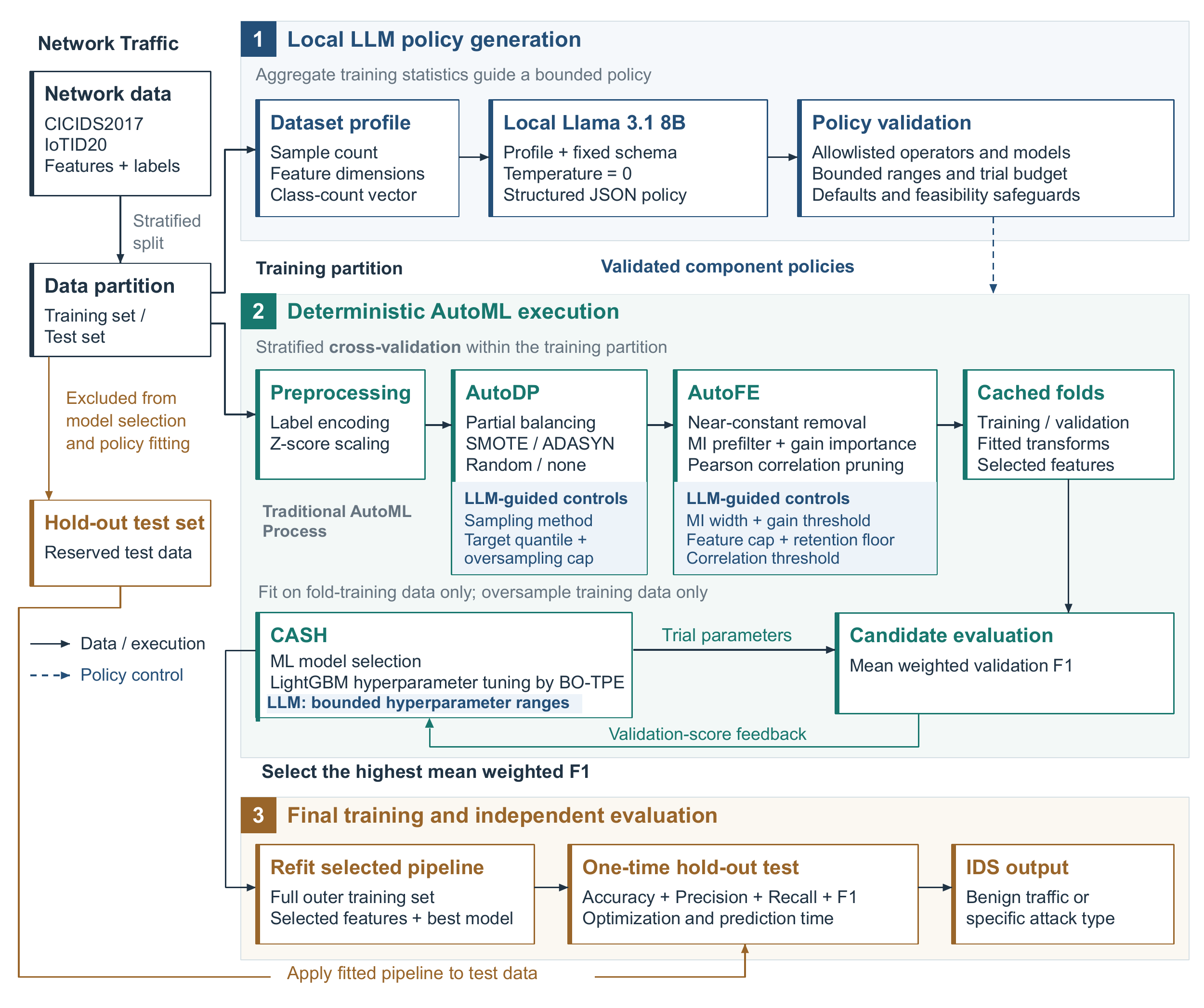}
 \caption{Overview of the proposed LLM-assisted AutoML IDS.} \label{framework}
\end{figure*}

The LLM policy is returned as JavaScript Object Notation (JSON). Within deterministic AutoML execution, AutoDP can use the Synthetic Minority Over-sampling Technique (SMOTE), Adaptive Synthetic Sampling (ADASYN), random oversampling, or no balancing. AutoFE combines near-constant feature removal, a Mutual Information (MI) prefilter, LightGBM gain-based importance selection, and Pearson correlation pruning. These operators and their policy-controlled parameters are shown explicitly in Fig.~\ref{framework}.

Fig.~\ref{framework} summarizes the workflow in three main stages. In Stage 1, the labeled traffic data are divided into an outer training partition and an independent hold-out set. Aggregate statistics from the training partition are provided to a local Llama 3.1 8B model \cite{llama3}, which returns a structured AutoML policy. The policy is validated against an admissible policy space, and unsupported choices are rejected or replaced by deterministic defaults. The hold-out set is excluded from policy fitting and model selection.

Stage 2 performs deterministic AutoML execution within stratified cross-validation on the outer training partition. First, label encoding and z-score scaling are fitted on each fold-training partition and applied to its paired validation fold. Second, AutoDP partially rebalances the fold-training data using the validated sampler, target class-count quantile, and per-class oversampling cap. Third, AutoFE applies near-constant removal, MI prefiltering, LightGBM gain-based selection, and Pearson correlation pruning with a final-feature retention safeguard. The fitted transformations and selected features are then cached with the paired training and validation folds. Fourth, CASH evaluates the validated LightGBM search policy with BO-TPE on these cached folds and selects the configuration with the highest mean weighted validation F1-score. In Stage 3, the selected pipeline is refitted on the full outer training set and evaluated once on the untouched hold-out set. Thus, the LLM provides bounded policy guidance, whereas data transformation, model fitting, and model selection remain deterministic or validation-driven. Table~\ref{tab:framework_components} summarizes the main components.

\begin{table}[tb]
\centering
\renewcommand{\arraystretch}{1.25}
\setlength{\extrarowheight}{1pt}
\setlength{\tabcolsep}{2.8pt}
\caption{Summary of the proposed LLM-assisted AutoML framework.}
\footnotesize
\begin{tabularx}{0.99\columnwidth}{|>{\centering\arraybackslash}p{1.25cm}|>{\raggedright\arraybackslash}p{2.95cm}|L|}
\hline
\textbf{Stage}
& \textbf{Objectives}
& \textbf{LLM-Assisted Decision} \\ \hline

Dataset profiling
& Compute sample count, feature types and dimensions, and class counts from the training partition.
& Provides aggregate evidence for AutoML policy generation. \\ \hline

Encoding and scaling
& Encode categorical predictors and scale numerical features.
& Uses a fixed process for consistency and does not participate in model scoring. \\ \hline

AutoDP balancing
& Partially rebalance minority classes within the training partition.
& Selects the sampler, target class-count quantile, and per-class oversampling cap. \\ \hline

AutoFE
& Remove weak, irrelevant, and redundant features while preserving a minimum feature set.
& Selects the mutual-information width, importance threshold, feature cap, retention floor, and correlation threshold. \\ \hline

CASH
& Optimize the validated model family and its hyperparameters.
& Provides bounded search-space guidance; the proposed branch uses a compact LightGBM-centered space and a small trial budget. \\ \hline

\end{tabularx}
\label{tab:framework_components}
\end{table}

Let $\mathcal{D}=\{(\mathbf{x}_i,y_i)\}_{i=1}^{n}$ denote a labeled traffic dataset, where $\mathbf{x}_i\in\mathbb{R}^{d}$ is a $d$-dimensional feature vector and $y_i\in\mathcal{C}$ is its class label. The proposed framework searches for an IDS classifier $f(\mathbf{x};\boldsymbol{\theta})$ that achieves high weighted F1-score on unseen traffic while limiting unnecessary AutoML evaluations. The LLM-assisted AutoML policy is represented as:
\begin{equation}
\boldsymbol{\pi}=\{\boldsymbol{\beta},\boldsymbol{\phi},\mathcal{A},\Omega,T\}\in\Pi,
\end{equation}
where $\boldsymbol{\beta}=\{b,q,r_{\max}\}$ contains the AutoDP controls: balancing method $b$, target class-count quantile $q$, and maximum per-class oversampling factor $r_{\max}$. The vector $\boldsymbol{\phi}$ denotes the feature-selection controls, $\mathcal{A}$ is the validated candidate model set, $\Omega$ is the corresponding conditional hyperparameter space, $T$ is the optimization budget, and $\Pi$ is the admissible policy space. 

The selected IDS model is obtained by maximizing the mean weighted F1-score under stratified cross-validation:
\begin{equation}
a^{*},\boldsymbol{\lambda}^{*}\in
\underset{a\in\mathcal{A},\boldsymbol{\lambda}\in\Omega_a}{\operatorname{argmax}}
\frac{1}{K}\sum_{k=1}^{K}F_1^{w}\left(a,\boldsymbol{\lambda},\mathcal{D}_{\mathrm{train}}^{(k)},\mathcal{D}_{\mathrm{valid}}^{(k)}\right),
\end{equation}
where $\mathcal{D}_{\mathrm{train}}^{(k)}$ and $\mathcal{D}_{\mathrm{valid}}^{(k)}$ are the training and validation folds in the $k$-th cross-validation split, respectively. Equation~(2) clarifies the role of the LLM: the LLM defines the AutoML policy, while the final model and its hyperparameters are selected according to measured validation performance.

\subsection{LLM Policy Generation and Validation}
The LLM-assisted stage begins by summarizing the outer training partition $\mathcal{D}_{\mathrm{train}}$ into a compact dataset profile $\mathcal{P}$. The profile includes the number of samples, feature dimensionality, numbers of numerical and categorical features, and the class-count vector. It is expressed as
\begin{equation}
\mathcal{P}(\mathcal{D}_{\mathrm{train}})=\{n,d,d_{\mathrm{num}},d_{\mathrm{cat}},\mathbf{c}_{y}\},
\quad
\mathbf{c}_{y}=\left[|\{i:y_i=c\}|\right]_{c\in\mathcal{C}},
\end{equation}
where $n$ is the number of samples, $d$ is the total number of input features, $d_{\mathrm{num}}$ and $d_{\mathrm{cat}}$ are the numbers of numerical and categorical features, respectively, $\mathbf{c}_{y}$ is the class-count vector, and $\mathcal{C}$ denotes the set of traffic classes. The class-count vector characterizes the degree of imbalance, while the dimensionality and feature-type counts provide evidence for feature-selection and search-space decisions. The profile contains aggregate information only; raw traffic records are not supplied to the LLM.

A local Llama~3.1 8B model is used as the policy generator \cite{llama3}. The model receives the aggregate profile together with a fixed JSON schema and is queried with temperature set to zero. No task-specific fine-tuning is performed. The prompt explicitly limits the LLM to policy generation: it does not access raw samples, score candidate models, or produce final IDS labels. The 8B variant was selected because it can be executed locally on the experimental workstation while following structured instructions. Local execution keeps the profile and experiment metadata within the experimental environment, avoids dependence on an external LLM service, and allows a fixed model version and decoding configuration to be used throughout the study.

The returned policy has three components: AutoDP, AutoFE, and CASH. AutoDP specifies the sampler together with the target class-count quantile and a per-class oversampling cap. AutoFE specifies the Pearson correlation threshold, cumulative gain-importance threshold, maximum retained-feature count, mutual-information prefilter width, and minimum final-feature count. CASH specifies candidate model information, a trial budget, and optional model-specific hyperparameter ranges. The raw JSON is never executed directly; it first passes through the validation layer described below.

The generated policy object is defined as
\begin{equation}
\mathcal{G}
=
\left\{
\begin{aligned}
\mathcal{G}_{\mathrm{AutoDP}} 
&= \{b,q,r_{\max}\}, \\[2pt]
\mathcal{G}_{\mathrm{AutoFE}} 
&= \left\{
\tau_{\mathrm{corr}},
\gamma_{\mathrm{imp}},
m_{\mathrm{imp}},
k_{\mathrm{MI}},
 m_{\min}
\right\}, \\[2pt]
\mathcal{G}_{\mathrm{CASH}} 
&= \left\{
\mathcal{A},
\Omega,
T
\right\}
\end{aligned}
\right\},
\end{equation}
where $b$, $q$, and $r_{\max}$ denote the balancing method, target quantile, and per-class oversampling cap, respectively. The AutoFE controls are the correlation threshold $\tau_{\mathrm{corr}}$, cumulative gain-importance threshold $\gamma_{\mathrm{imp}}$, feature cap $m_{\mathrm{imp}}$, mutual-information width $k_{\mathrm{MI}}$, and minimum final-feature count $m_{\min}$. The variables $\mathcal{A}$, $\Omega$, and $T$ denote the validated candidate set, conditional hyperparameter space, and optimization budget.

The validation layer converts the JSON response into an executable policy by applying an allowlist and numeric bounds. The balancing method is restricted to SMOTE, ADASYN, random oversampling, or no balancing \cite{smote,adasyn}; the target quantile and oversampling factor are clipped to bounded ranges. AutoFE thresholds and feature counts are likewise constrained, including the minimum final-feature count. Model names are limited to LightGBM, random forest, and extra trees, and invalid or missing fields are replaced by deterministic defaults. For the proposed low-budget CASH branch, the executable policy uses LightGBM as the model family and a compact predefined search envelope. If the LLM provides model-specific LightGBM ranges, their intersection with this envelope is used; an invalid or missing intersection leaves the predefined range unchanged.

Additional feasibility checks are applied during execution. The number of neighbors for SMOTE or ADASYN is reduced when a sampled class is small, and random oversampling is used if synthetic sampling cannot be performed. AutoFE also enforces feature-retention safeguards before and after correlation pruning. These rules prevent malformed LLM output or small-class geometry from changing the experimental protocol.

\subsection{LLM-Assisted Automated Data Balancing}

Before data balancing and model learning, categorical traffic attributes are converted into numerical values using label encoding \cite{sklearn}. For each categorical feature, an encoder is fitted on the active training partition and then applied to the paired validation or hold-out partition. Previously unseen categories are mapped to a dedicated unknown token. This transformation allows tree-based models and feature selectors to process categorical fields such as addresses, ports, or protocol identifiers without expanding the feature space through one-hot encoding. After encoding, all numerical features are normalized by z-score scaling. For a feature value $x$, the normalized value is \cite{sklearn}:
\begin{equation}
x^{\prime}=\frac{x-\mu_{\mathrm{tr}}}{\sigma_{\mathrm{tr}}},
\end{equation}
where $\mu_{\mathrm{tr}}$ and $\sigma_{\mathrm{tr}}$ are the mean and standard deviation estimated from the active training partition. The same parameters are then applied to the paired validation or hold-out data. Numerical features are standardized before balancing so that distance-based oversampling is not dominated by features with larger numerical ranges. Applying the same label encoding and z-score normalization protocol to all comparison methods ensures that the observed differences come from balancing, feature selection, and CASH policies rather than inconsistent pre-processing.

Network intrusion datasets are often highly imbalanced because normal traffic and frequent attack categories may dominate the data, whereas rare attacks contain far fewer samples \cite{iot_evcs_survey}. The AutoDP module computes the class-count vector $\mathbf{c}$ on the active training partition and the imbalance ratio
\begin{equation}
\rho=\frac{\max(\mathbf{c})}{\min(\mathbf{c})}.
\end{equation}
No resampling is performed when $\rho\leq1.5$. Otherwise, AutoDP performs partial balancing rather than forcing every minority class to the majority count. For a validated target quantile $q$, the common target is
\begin{equation}
t_q=\min\left(\max(\mathbf{c}),\max\left(Q_q(\mathbf{c}),\min(\mathbf{c})+1\right)\right),
\end{equation}
where $Q_q(\mathbf{c})$ is the $q$-quantile of the class-count distribution. In the conventional AutoDP setting, $q=0.75$ and no per-class growth cap is applied. In the LLM-assisted setting, the policy also provides $r_{\max}$; therefore, a class $j$ with count $c_j$ is assigned
\begin{equation}
t_j=\min\left(t_q,\left\lfloor r_{\max}c_j\right\rfloor\right),
\end{equation}
and it is oversampled only when $c_j<t_j$. This cap is particularly useful for very small classes because it avoids a large increase in the training-set size.

The policy chooses among SMOTE, ADASYN, random oversampling, and no balancing. SMOTE interpolates between neighboring minority samples \cite{smote}, whereas ADASYN allocates more synthetic samples to locally difficult minority regions \cite{adasyn}. For SMOTE and ADASYN, the neighborhood size is reduced when a sampled class contains only a few examples. If fewer than two samples are available, or if synthetic sampling otherwise fails, random oversampling is used as a deterministic fallback. All resampling is fitted on fold-training data only; validation and hold-out samples are never used to construct synthetic data.

\subsection{LLM-Assisted Automated Feature Engineering}

Feature engineering aims to improve the feature set used by ML models. In network intrusion detection, raw datasets often contain weakly informative, redundant, or near-constant features. Feature selection, a major component of feature engineering, removes unnecessary features while retaining those important for intrusion detection \cite{automl_iot,ids1}. This step is important for IoT IDS development because fewer features can reduce data complexity, model execution time, and the risk of overfitting.

The proposed AutoFE process has four stages: near-constant feature removal, mutual-information prefiltering, LightGBM gain-based cumulative importance selection, and Pearson correlation pruning. All selectors are fitted on the active training partition only. The LLM controls the mutual-information width, cumulative importance threshold, feature cap, minimum final-feature count, and correlation threshold. The stages serve different purposes: variance filtering removes nearly constant inputs, mutual information provides a fast relevance screen, LightGBM gain importance provides model-aware ranking, and correlation pruning removes redundant retained features.

The first step removes near-constant features using a variance threshold of $10^{-8}$. Let $\mathcal{F}_{v}$ denote the feature set remaining after variance filtering, and let $d_{v}=|\mathcal{F}_{v}|$. The second step applies a mutual-information prefilter. Let $k_{\mathrm{MI}}$ denote the LLM-selected mutual-information width and $m_{\mathrm{imp}}$ denote the maximum number of important features. The prefilter width is defined as:
\begin{equation}
m_{\mathrm{MI}}
=
\min
\left\{
\max
\left(
3m_{\mathrm{imp}},
k_{\mathrm{MI}},
60
\right),
d_{v}
\right\}.
\end{equation}
If $d_{v}>m_{\mathrm{MI}}$, only the top $m_{\mathrm{MI}}$ features ranked by mutual information with the class label are retained.

The third step trains a LightGBM classifier on the training partition and ranks features using gain-based importance. Let $q_{j}$ denote the raw LightGBM gain importance of feature $j$. The normalized importance score is defined as \cite{automl_ccs}:
\begin{equation}
I_{j}
=
\frac{q_{j}}
{\sum_{r\in\mathcal{F}_{\mathrm{MI}}} q_{r}},
\qquad
j\in\mathcal{F}_{\mathrm{MI}},
\end{equation}
where $\mathcal{F}_{\mathrm{MI}}$ is the feature set retained after mutual-information prefiltering. Features are sorted in descending order of $I_{j}$, and the smallest subset whose cumulative importance reaches the LLM-selected threshold $\gamma_{\mathrm{imp}}$ is retained:
\begin{equation}
\left\{
\begin{aligned}
\mathcal{F}_{\mathrm{imp}}
&=
\left\{
j_{1},j_{2},\ldots,j_{m^{*}}
\right\}, \\[2pt]
m^{*}
&=
\min
\left\{
m:
\sum_{\ell=1}^{m} I_{j_{\ell}}
\geq
\gamma_{\mathrm{imp}}
\right\}.
\end{aligned}
\right.
\end{equation}
where $I_{j_{1}}\geq I_{j_{2}}\geq\cdots$ after sorting. In implementation, the cumulative threshold is evaluated with a single LightGBM fit, and the retained count is also bounded by $m_{\mathrm{imp}}$. To avoid an overly aggressive reduction, the importance stage retains at least
\begin{equation}
m_{\mathrm{floor}}^{\mathrm{imp}}=
\min\left(m_{\mathrm{imp}},\max\left(30,\left\lceil0.8m_{\mathrm{imp}}\right\rceil,m_{\min}\right)\right)
\end{equation}
features when the importance selector is active. If the MI-reduced feature set already contains no more than the requested feature cap, the importance-selection step is skipped. This single-fit selector avoids the repeated model fitting required by recursive elimination.

The final step removes redundant features using the absolute Pearson correlation matrix estimated from the training partition. For two selected features $p$ and $q$, the absolute correlation is computed as
\begin{equation}
R_{pq}
=
\left|
\operatorname{corr}
\left(
\mathbf{x}_{p},
\mathbf{x}_{q}
\right)
\right|.
\end{equation}
If $R_{pq}>\tau_{\mathrm{corr}}$, one of the two features is removed. The LLM-assisted policy includes a retention floor $m_{\min}$. If correlation pruning leaves fewer than $m_{\min}$ features, features are restored from the pre-correlation subset until the floor is met; if no feature remains, the pre-correlation subset is retained. This safeguard is applied after the MI and gain-importance stages so that correlation pruning cannot collapse an already compact feature set.

Thus, the LLM-assisted AutoFE policy adapts five controls to the dataset profile: $k_{\mathrm{MI}}$, $\gamma_{\mathrm{imp}}$, $m_{\mathrm{imp}}$, $m_{\min}$, and $\tau_{\mathrm{corr}}$. The deterministic bounds are intentionally conservative for high-dimensional data, so the LLM can reduce redundant inputs without selecting an impractically small final feature set.

\subsection{LLM-Assisted Automated Model Learning and CASH Optimization}
After pre-processing, balancing, and AutoFE, the framework performs BO-TPE-based model optimization. LightGBM, Random Forest (RF), and Extra Trees (ET) are retained in the conventional ML and AutoML comparisons because they are strong and efficient models for tabular network-flow data. The proposed low-budget LLM-assisted CASH branch uses LightGBM as its executable model family and optimizes it within a compact validated hyperparameter space. This keeps the model family fixed across LLM-assisted trials while allowing the LLM policy to guide the search bounds and budget.

RF constructs an ensemble of decision trees trained on bootstrapped samples and random subsets of features \cite{iot_idsai}. Its prediction is obtained by aggregating the votes of individual trees. RF is robust, relatively insensitive to feature scaling, and effective for high-dimensional tabular data. It is used as a strong plain ML and standard AutoML baseline because it often performs well on IDS datasets with limited tuning.

ET further randomizes both feature selection and split thresholds \cite{iot_idsai}. This stronger randomization can reduce variance and training cost, but it may require a sufficiently large ensemble to achieve the best performance. In the experiments, ET provides an efficiency-oriented baseline and helps evaluate whether the additional randomness improves or degrades IDS classification under the selected datasets.

LightGBM \cite{lightgbm} is a gradient-boosted decision-tree algorithm designed for efficient large-scale learning. It grows boosted trees sequentially, where each new tree corrects the errors of the current ensemble. LightGBM improves efficiency through histogram-based binning, Gradient-based One-Side Sampling (GOSS), and Exclusive Feature Bundling (EFB). These properties make it especially suitable for tabular network traffic with many numerical features. In the proposed LLM-assisted branch, LightGBM is selected as the final model family because it provides a favorable balance between detection performance and search efficiency.

The learning objective of a gradient-boosted tree ensemble can be written as \cite{lightgbm}:
\begin{equation}
\hat{y}_i=\sum_{m=1}^{M} f_m(\mathbf{x}_i), \quad f_m \in \mathcal{F},
\end{equation}
\begin{equation}
\mathcal{L}=\sum_{i=1}^{n} \ell(y_i,\hat{y}_i)+\sum_{m=1}^{M}\mathcal{R}(f_m).
\end{equation}
Here, $f_m$ is an individual tree, $\ell(\cdot)$ is the classification loss, and $\mathcal{R}(\cdot)$ is a regularization term that penalizes overly complex trees. In LightGBM, important hyperparameters include the number of estimators, maximum tree depth, learning rate, number of leaves, minimum child samples, subsampling ratio, feature subsampling ratio, and L1/L2 regularization terms. In RF and ET, important hyperparameters include the number of trees, maximum depth, minimum split size, minimum leaf size, and maximum feature ratio.

BO-TPE is selected as the HPO method because it is well suited to mixed and conditional search spaces \cite{tpe}. Unlike grid search, BO-TPE does not exhaustively evaluate every combination. Unlike random search, it uses previous observations to guide future trials. Compared with Gaussian-process Bayesian optimization, TPE handles categorical, integer, and conditional parameters more naturally. This is important for CASH because each model family has a different hyperparameter space.

TPE models the density of good and poor configurations separately. Let $y$ be the validation loss or negative validation score, and let $y^{*}$ be a quantile threshold that separates promising configurations from weaker ones. TPE estimates
\begin{equation}
p(\boldsymbol{\lambda}|y)=\begin{cases} l(\boldsymbol{\lambda}), & y<y^{*}, \\ g(\boldsymbol{\lambda}), & y\geq y^{*} \end{cases}.
\end{equation}
New configurations are sampled to increase the ratio $l(\boldsymbol{\lambda})/g(\boldsymbol{\lambda})$, which favors regions that are more likely under good trials than under poor trials. In this work, Optuna implements the BO-TPE optimizer \cite{optuna}. The objective function is the mean weighted F1-score under stratified cross-validation.
\begin{equation}
J(a,\boldsymbol{\lambda})=\frac{1}{K}\sum_{k=1}^{K}F_1^w(a,\boldsymbol{\lambda},\mathcal{D}_{tr}^{(k)},\mathcal{D}_{val}^{(k)}).
\end{equation}

The LLM-assisted CASH stage is configured before BO-TPE begins. The prompt allows bounded model-specific ranges, but the executable policy fixes LightGBM for the proposed branch to keep the optimization cost stable. A predefined compact LightGBM envelope provides the safe search space; when the LLM returns valid LightGBM bounds, they are intersected with that envelope, and missing or invalid bounds leave the predefined ranges unchanged. The proposed branch uses a 10-trial budget, whereas the standard CASH baseline searches LightGBM, RF, and ET with 30 trials. BO-TPE and cross-validation, rather than the LLM, determine the final hyperparameters from measured validation F1-score.

\begin{algorithm}[t]
\caption{Proposed LLM-Assisted AutoML IDS Framework}
\label{alg:llm_automl_ids}
\footnotesize
\begin{algorithmic}[1]

\STATE \textbf{Input:} Labeled dataset $\mathcal{D}=\{X,Y\}$; local LLM $\mathcal{M}_{\mathrm{LLM}}$; admissible policy space $\Pi$; number of CV folds $K$.
\STATE \textbf{Output:} Optimized IDS $\mathcal{M}^{*}$; validated policy $\boldsymbol{\pi}$; selected features $\mathcal{F}^{*}$; best configuration $(a^{*},\boldsymbol{\lambda}^{*})$; evaluation results $\mathcal{R}$.

\STATE \textit{// Step 1: LLM-assisted AutoML policy generation and validation}
\STATE Split $\mathcal{D}\rightarrow\{\mathcal{D}_{\mathrm{train}},\mathcal{D}_{\mathrm{test}}\}$ using a stratified outer split; reserve $\mathcal{D}_{\mathrm{test}}$ for final evaluation.
\STATE $\mathcal{P}\leftarrow\{n,d,d_{\mathrm{num}},d_{\mathrm{cat}},\mathbf{c}_{y}\}$ from $\mathcal{D}_{\mathrm{train}}$ only.
\STATE $\mathcal{G}\leftarrow\mathcal{M}_{\mathrm{LLM}}(\mathcal{P})$.
\COMMENT{LLM proposes bounded AutoDP, AutoFE, and CASH settings}
\STATE $\boldsymbol{\pi}\leftarrow\operatorname{Validate}(\mathcal{G},\Pi)$.
\COMMENT{Deterministic validation corrects invalid or missing settings}
\STATE Extract $\boldsymbol{\beta}=\{b,q,r_{\max}\}$, $\boldsymbol{\phi}=\{k_{\mathrm{MI}},\gamma_{\mathrm{imp}},m_{\mathrm{imp}},\tau_{\mathrm{corr}},m_{\min}\}$, and $\{\mathcal{A},\Omega,T\}$ from $\boldsymbol{\pi}$.

\STATE \textit{// Step 2: LLM-guided, deterministic AutoDP and AutoFE}
\STATE Create $K$ stratified folds $\{(\mathcal{D}_{\mathrm{tr}}^{(k)},\mathcal{D}_{\mathrm{val}}^{(k)})\}_{k=1}^{K}$ from $\mathcal{D}_{\mathrm{train}}$.
\FOR{$k=1,\ldots,K$}
    \STATE Fit label encoding and z-score scaling on $\mathcal{D}_{\mathrm{tr}}^{(k)}$; transform both fold partitions.
    \COMMENT{Fixed preprocessing; no LLM decision}
    
    \STATE $\rho^{(k)}\leftarrow \max_j c_j^{(k)}/\min_j c_j^{(k)}$.
    \IF{$\rho^{(k)}>1.5$ and $b\neq\mathrm{None}$}
        \STATE $\mathbf{t}^{(k)}\leftarrow\operatorname{Targets}(\mathbf{c}^{(k)},q,r_{\max})$.
        \STATE $\mathcal{D}_{\mathrm{tr}}^{(k)}\leftarrow
        \operatorname{AutoDP}(\mathcal{D}_{\mathrm{tr}}^{(k)},b,\mathbf{t}^{(k)})$.
        \COMMENT{LLM selects bounded balancing policy; operator performs resampling}
        \STATE Adjust neighbor count when needed; use random oversampling if SMOTE/ADASYN is infeasible.
    \ENDIF
    
    \STATE $\mathcal{F}_{v}^{(k)}\leftarrow\operatorname{VarianceFilter}
    (\mathcal{D}_{\mathrm{tr}}^{(k)})$.
    \STATE $\mathcal{F}_{\mathrm{MI}}^{(k)}\leftarrow
    \operatorname{TopMI}(\mathcal{F}_{v}^{(k)},k_{\mathrm{MI}})$.
    \STATE $\mathcal{F}_{\mathrm{imp}}^{(k)}\leftarrow
    \operatorname{GainSelect}(\mathcal{F}_{\mathrm{MI}}^{(k)},
    \gamma_{\mathrm{imp}},m_{\mathrm{imp}})$.
    \STATE $\mathcal{F}^{(k)}\leftarrow
    \operatorname{CorrPrune}(\mathcal{F}_{\mathrm{imp}}^{(k)},
    \tau_{\mathrm{corr}},m_{\min})$.
    \COMMENT{LLM sets AutoFE controls; feature selection is fitted on training data}
    
    \STATE Apply $\mathcal{F}^{(k)}$ to the training and validation partitions and cache the prepared fold $\mathcal{V}^{(k)}$.
\ENDFOR

\STATE \textit{// Step 3: LLM-constrained CASH optimization using BO-TPE}
\STATE Initialize BO-TPE over $\{(a,\Omega_a):a\in\mathcal{A}\}$ with trial budget $T$.
\COMMENT{LLM defines the bounded search policy; BO-TPE performs empirical search}
\FOR{$t=1,\ldots,T$}
    \STATE $(a_t,\boldsymbol{\lambda}_t)\leftarrow
    \operatorname{TPE}(\mathcal{A},\Omega)$.
    \FOR{$k=1,\ldots,K$}
        \STATE Train $(a_t,\boldsymbol{\lambda}_t)$ on the cached training data in $\mathcal{V}^{(k)}$.
        \STATE $s_t^{(k)}\leftarrow
        F_1^{w}(a_t,\boldsymbol{\lambda}_t;\mathcal{V}^{(k)})$ on the paired validation data.
    \ENDFOR
    \STATE $\bar{s}_t\leftarrow\frac{1}{K}\sum_{k=1}^{K}s_t^{(k)}$; update BO-TPE using $\bar{s}_t$.
\ENDFOR
\STATE $(a^{*},\boldsymbol{\lambda}^{*})
\leftarrow\operatorname*{arg\,max}_{t=1,\ldots,T}\bar{s}_t$.
\COMMENT{Final configuration is selected by measured CV performance, not by the LLM}

\STATE \textit{// Step 4: Final pipeline fitting and independent evaluation}
\STATE Fit encoding and z-score scaling on $\mathcal{D}_{\mathrm{train}}$ and apply them to both outer partitions.
\STATE Apply $\operatorname{AutoDP}(\cdot;\boldsymbol{\beta})$ to $\mathcal{D}_{\mathrm{train}}$ only.
\STATE $\mathcal{F}^{*}\leftarrow
\operatorname{AutoFE}(\mathcal{D}_{\mathrm{train}};\boldsymbol{\phi})$; apply $\mathcal{F}^{*}$ to both outer partitions.
\STATE $\mathcal{M}^{*}\leftarrow
\operatorname{Train}(a^{*},\boldsymbol{\lambda}^{*},
\mathcal{D}_{\mathrm{train}},\mathcal{F}^{*})$.
\STATE $\mathcal{R}\leftarrow
\operatorname{Evaluate}(\mathcal{M}^{*},\mathcal{D}_{\mathrm{test}})$.
\COMMENT{LLM has no role in model scoring or hold-out evaluation}

\STATE \textbf{return} $\mathcal{M}^{*}$, $\boldsymbol{\pi}$, $\mathcal{F}^{*}$, $(a^{*},\boldsymbol{\lambda}^{*})$, and $\mathcal{R}$.

\end{algorithmic}
\end{algorithm}

\subsection{Proposed Framework Summary and Deployment}

Algorithm \ref{alg:llm_automl_ids} summarizes the complete training and evaluation procedure. The framework first creates an independent outer train--test split, constructs the dataset profile from the training partition, and converts the local LLM output into a validated policy for AutoDP, AutoFE, and CASH. The deterministic AutoML stage then prepares the stratified cross-validation folds by fitting encoding and scaling, applying partial AutoDP balancing to fold-training data only, and fitting the four-stage AutoFE process before caching the prepared folds. BO-TPE evaluates the validated CASH space on these cached folds and selects the configuration with the highest mean weighted validation F1-score. Finally, the selected pipeline is refitted on the full outer training partition, and the hold-out test set is used once for independent evaluation.

The proposed framework provides four main advantages:
\begin{enumerate}
    \item It improves search efficiency by adapting a compact AutoML policy to the dataset profile rather than exploring a broad, generic search space.
    \item It supports reproducible execution because each generated policy is validated against predefined ranges and corrected using deterministic defaults when necessary.
    \item It supports local cybersecurity applications because Llama~3.1 8B can operate without transmitting experimental information to an external LLM service.
    \item It preserves experimental rigor by limiting the LLM to policy generation, while model selection remains based on cross-validation and the hold-out test set is reserved for final evaluation.
\end{enumerate}

Traditional CASH baselines spend most of their runtime repeatedly training models across a broad candidate space. In contrast, the proposed method performs policy generation and validation once, followed by fewer BO-TPE trials within a narrower conditional space~\cite{optuna}. This shifts the computational effort from broad exploration to guided optimization, while cross-validation remains responsible for selecting the final model.

The framework is also suitable for constrained IoT environments. Model training and LLM-assisted planning can be performed on a gateway, server, or cloud platform, while the optimized tree-based IDS can be deployed close to the traffic source. Tree ensembles provide efficient prediction for tabular network traffic~\cite{tabular_tree,lightgbm}, and the LLM is required only during AutoML planning rather than during IDS inference. This deployment model is deliberately different from direct LLM-based IDS agents~\cite{ids_agent,huntgpt}: after model development, the LLM is not part of the detection loop and does not inspect live traffic or generate operational alerts.

A further practical advantage is that the generated policy can be cached and audited. The stored policy records the selected balancing method, AutoFE thresholds, candidate model family, hyperparameter ranges, and trial budget. This makes the AutoML configuration easier to reproduce than an iterative manual tuning process and allows researchers to inspect the LLM-generated decisions before expensive model search is executed. Such inspection is useful in cybersecurity applications, where an unreasonable policy can be identified before it affects the training pipeline.

Overall, the framework uses the LLM as a bounded source of prior knowledge while leaving model fitting, selection, and evaluation to the validated AutoML pipeline. This separation allows the LLM to guide a more focused search without becoming part of the deployed detector, while keeping the development process efficient, inspectable, and reproducible.

\section{Performance Evaluation} \label{S4}
\subsection{Experimental Setup}
All experiments in this work were conducted on a Lenovo Legion 5 machine with an Intel Core Ultra 7 255HX Central Processing Unit (CPU), an NVIDIA RTX 5070 Graphics Processing Unit (GPU), and 32 GB of Random-Access Memory (RAM), representing an IoT server machine. The experiments use Python-based ML libraries, including Scikit-learn for classical ML procedures, Imbalanced-learn \cite{imblearn} for oversampling, LightGBM for LightGBM model development, Optuna for BO-TPE optimization, and Ollama with Llama~3.1 8B for local LLM-based AutoML policy generation. A fixed random seed of 0 is used for train-test splitting, stratified cross-validation, model construction, sampling, and HPO. This improves reproducibility across the main comparison and ablation studies.

The proposed framework was evaluated on two public benchmark intrusion detection datasets, CICIDS2017 \cite{cicids2017} and IoTID20 \cite{iotid20}. CICIDS2017 is a widely used benchmark cybersecurity dataset that contains modern network traffic with benign samples and multiple attack types such as DoS, brute force, web attacks, botnet, port scan, and infiltration attacks. IoTID20 is an IoT-focused benchmark dataset containing benign traffic and several IoT attack categories, including Mirai, DoS, scanning, and MITM attacks. These two datasets are selected because they jointly represent general network and IoT-specific intrusion detection scenarios, and both contain various types of attacks, heterogeneous network-flow features, and highly imbalanced class distributions.

Given the focus of this work on resource-aware IDS development for IoT and edge environments, representative subsets of CICIDS2017 and IoTID20 were used for model development and evaluation. The use of representative subsets reduces the computational and memory burden of repeated fold-wise preprocessing, AutoFE, and CASH optimization, while retaining the evaluated normal/benign traffic and attack categories as well as their pronounced class imbalance. This is also consistent with practical IoT settings, where gateways and edge systems may have limited storage and computing resources and may not retain or repeatedly process all historical traffic records \cite{pwpae,msana_iiot}. The same fixed subsets are used by all compared methods to maintain a consistent evaluation protocol. The final CICIDS2017 subset contains 26,800 samples, while the IoTID20 subset contains 62,578 samples.

\begin{table*}[tb]
\centering
\renewcommand{\arraystretch}{1.25}
\setlength{\extrarowheight}{1pt}
\setlength{\tabcolsep}{4.5pt}
\caption{Class distributions of the representative CICIDS2017 and IoTID20 subsets used in the experiments.}
\footnotesize
\scalebox{0.82}{
\begin{tabular}{|C{2.0cm}|C{3.2cm}|C{2.1cm}|C{2.1cm}|C{2.5cm}|C{2.1cm}|}
\hline
\textbf{Dataset}
& \textbf{Class}
& \textbf{Samples}
& \textbf{Distribution (\%)}
& \textbf{Training samples for Cross Validation}
& \textbf{Hold-Out Test samples} \\ \hline

\multirow{8}{*}{CICIDS2017}
& Benign       & 18,225 & 68.004 & 14,580 & 3,645 \\ \cline{2-6}
& DoS          & 3,042  & 11.351 & 2,433  & 609   \\ \cline{2-6}
& Web Attack   & 2,180  & 8.134  & 1,744  & 436   \\ \cline{2-6}
& Botnet       & 1,966  & 7.336  & 1,573  & 393   \\ \cline{2-6}
& Port Scan    & 1,255  & 4.683  & 1,004  & 251   \\ \cline{2-6}
& Brute Force  & 96     & 0.358  & 77     & 19    \\ \cline{2-6}
& Infiltration & 36     & 0.134  & 29     & 7     \\ \cline{2-6}
& \textbf{Total} & \textbf{26,800} & \textbf{100.000} & \textbf{21,440} & \textbf{5,360} \\ \hline

\multirow{6}{*}{IoTID20}
& Mirai              & 41,340 & 66.062 & 33,072 & 8,268 \\ \cline{2-6}
& Scan               & 7,620  & 12.177 & 6,096  & 1,524 \\ \cline{2-6}
& DoS                & 6,047  & 9.663  & 4,837  & 1,210 \\ \cline{2-6}
& Normal             & 4,031  & 6.442  & 3,225  & 806 \\ \cline{2-6}
& MITM ARP Spoofing  & 3,540  & 5.657  & 2,832  & 708 \\ \cline{2-6}
& \textbf{Total} & \textbf{62,578} & \textbf{100.000} & \textbf{50,062} & \textbf{12,516} \\ \hline
\end{tabular}
}
\label{tab:dataset_distribution}
\end{table*}

Table~\ref{tab:dataset_distribution} shows that both evaluated subsets remain highly imbalanced, which is important for evaluating the AutoDP component. In CICIDS2017, benign traffic accounts for 68.004\% of the subset, whereas infiltration and brute-force samples represent only 0.134\% and 0.358\%, respectively. In IoTID20, Mirai traffic accounts for 66.062\% of the subset, while the other classes are substantially smaller. These distributions provide a useful setting for examining whether the proposed partial-balancing policy can improve minority-class representation without unnecessarily expanding every class to the majority size.

Each dataset is divided into an outer training set and a hold-out test set using an 80/20 stratified split, with the resulting sample counts also reported in Table~\ref{tab:dataset_distribution}. Model selection and HPO use the outer training set only. To keep repeated CASH evaluation computationally manageable, a fixed stratified 12,000-sample search subset is drawn from the outer training partition and used for 3-fold stratified cross-validation. The same search-subset rule is used for the plain and AutoML search comparisons, and the hold-out set is never used for this sampling or for model selection. Within each fold, label encoding and z-score normalization are fitted on the fold-training data, AutoDP resamples only the fold-training data under its partial-balancing policy, and AutoFE is fitted on the resulting training fold and then applied to the paired validation fold. After HPO, the selected pipeline is fitted on the complete outer training set and evaluated once on the hold-out test set.

Accuracy and weighted precision, recall, and F1-score are used to evaluate the proposed IDSs comprehensively. The F1-score balances precision and recall and is particularly useful when false positives and false negatives both matter. Efficiency is evaluated using optimizer time and hold-out prediction time per sample. Optimizer time measures the model-search procedure itself; the one-time LLM policy-generation latency is tracked separately and included in the total wall-clock time rather than in the optimizer-time column. The LLM calls used in the reported experiments required 6.88~s for CICIDS2017 and 4.99~s for IoTID20, produced valid JSON outputs, and did not trigger a fallback. The conventional CASH comparisons use Random Search (RS) and BO-TPE as their optimization strategies.

\subsection{Experimental Results and Discussion}

\begin{table*}[tb]
\centering
\renewcommand{\arraystretch}{1.30}
\setlength{\extrarowheight}{1pt}
\setlength{\tabcolsep}{3.5pt}
\caption{Classification performance and efficiency comparison on CICIDS2017 and IoTID20.}
\scalebox{0.79}{
\begin{tabular}{|C{1.7cm}|C{3.0cm}|C{2.4cm}|C{1.4cm}|C{1.4cm}|C{1.4cm}|C{1.4cm}|C{1.6cm}|C{1.9cm}|C{1.2cm}|}
\hline
\textbf{Dataset} 
& \textbf{Method} 
& \textbf{Best ML model} 
& \textbf{Test Accuracy (\%)} 
& \textbf{Test Precision (\%)} 
& \textbf{Test Recall (\%)} 
& \textbf{Test F1 (\%)} 
& \textbf{Optimization time (s)} 
& \textbf{Test time per sample (ms)} 
& \textbf{Features} \\ \hline

\multirow{11}{*}{CICIDS2017}
& PSO-LSTM \cite{pso_lstm_ids} & LSTM & 95.951 & 95.997 & 95.951 & 95.956 & 1110.95 & 0.0629 & 23 \\ \cline{2-10}
& GA-CNN \cite{ga_cnn_ids} & CNN & 96.679 & 96.809 & 96.679 & 96.704 & 960.11 & 0.0703 & 77 \\ \cline{2-10}
& OE-IDS \cite{oe_ids} & Soft Voting & 99.384 & 99.387 & 99.384 & 99.378 & 382.67 & 0.0582 & 15 \\ \cline{2-10}
& AutoML-ID \cite{automl_id} & Boosting Ensemble & 99.515 & 99.516 & 99.515 & 99.515 & 292.91 & 0.0124 & 77 \\ \cline{2-10}
& LightGBM \cite{lightgbm_ids} & LightGBM & 99.422 & 99.430 & 99.422 & 99.420 & 3.60 & 0.0053 & 77 \\ \cline{2-10}
& RF \cite{iot_idsai} & Random Forest & 99.571 & 99.578 & 99.571 & 99.569 & 2.35 & 0.0091 & 77 \\ \cline{2-10}
& ET \cite{iot_idsai} & Extra Trees & 98.433 & 98.514 & 98.433 & 98.450 & 1.11 & 0.0109 & 77 \\ \cline{2-10}
& Traditional AutoML-RS & LightGBM & 99.515 & 99.523 & 99.515 & 99.514 & 111.80 & 0.0079 & 49 \\ \cline{2-10}
& Traditional AutoML-TPE - 10 trials \cite{automl_ccs} & LightGBM & 99.534 & 99.539 & 99.534 & 99.532 & 37.41 & 0.0145 & 49 \\ \cline{2-10}
& Traditional AutoML-TPE - 30 trials \cite{automl_ccs} & LightGBM & 99.627 & 99.630 & 99.627 & 99.624 & 128.41 & 0.0208 & 49 \\ \cline{2-10}
& \textbf{Proposed LLM-AutoML - 10 trials} & LightGBM & \textbf{99.683} & \textbf{99.685} & \textbf{99.683} & \textbf{99.680} & 46.55 & 0.0198 & 41 \\ \hline

\multirow{11}{*}{IoTID20}
& PSO-LSTM \cite{pso_lstm_ids} & LSTM & 95.262 & 95.228 & 95.262 & 95.121 & 1281.44 & 0.0203 & 28 \\ \cline{2-10}
& GA-CNN \cite{ga_cnn_ids} & CNN & 98.154 & 98.167 & 98.154 & 98.155 & 1480.58 & 0.0401 & 81 \\ \cline{2-10}
& OE-IDS \cite{oe_ids} & Soft Voting & 97.643 & 97.649 & 97.643 & 97.616 & 294.70 & 0.0266 & 19 \\ \cline{2-10}
& AutoML-ID \cite{automl_id} & Boosting Ensemble & 99.017 & 99.022 & 99.017 & 99.019 & 239.80 & 0.0403 & 81 \\ \cline{2-10}
& LightGBM \cite{lightgbm_ids} & LightGBM & 98.778 & 98.809 & 98.778 & 98.787 & 5.12 & 0.0070 & 81 \\ \cline{2-10}
& RF \cite{iot_idsai} & Random Forest & 99.073 & 99.080 & 99.073 & 99.075 & 1.94 & 0.0050 & 81 \\ \cline{2-10}
& ET \cite{iot_idsai} & Extra Trees & 98.634 & 98.670 & 98.634 & 98.645 & 1.32 & 0.0066 & 81 \\ \cline{2-10}
& Traditional AutoML-RS & Random Forest & 99.081 & 99.101 & 99.081 & 99.087 & 107.25 & 0.0076 & 59 \\ \cline{2-10}
& Traditional AutoML-TPE - 10 trials \cite{automl_ccs} & LightGBM & 98.834 & 98.869 & 98.834 & 98.844 & 29.66 & 0.0149 & 59 \\ \cline{2-10}
& Traditional AutoML-TPE - 30 trials \cite{automl_ccs} & Random Forest & 99.169 & 99.173 & 99.169 & 99.170 & 72.36 & 0.0059 & 59 \\ \cline{2-10}
& \textbf{Proposed LLM-AutoML - 10 trials} & LightGBM & \textbf{99.185} & \textbf{99.187} & \textbf{99.185} & \textbf{99.186} & 36.28 & 0.0196 & 19 \\ \hline

\end{tabular}
}
\label{tab:performance_efficiency}
\end{table*}

\begin{figure*}[tb]
 \centering
 \includegraphics[width=17.5cm]{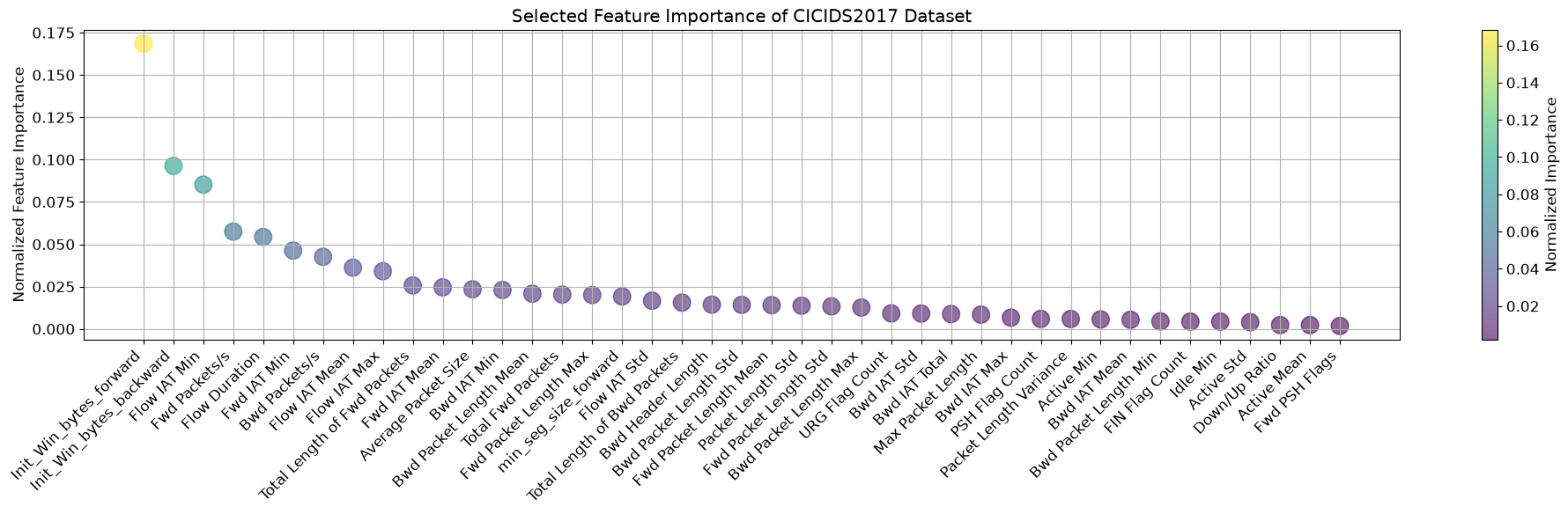}
 \caption{Selected features and normalized importance scores of the final LLM-assisted IDS on CICIDS2017.} \label{fs_cic}
\end{figure*}

\begin{figure*}[tb]
 \centering
 \includegraphics[width=17.5cm]{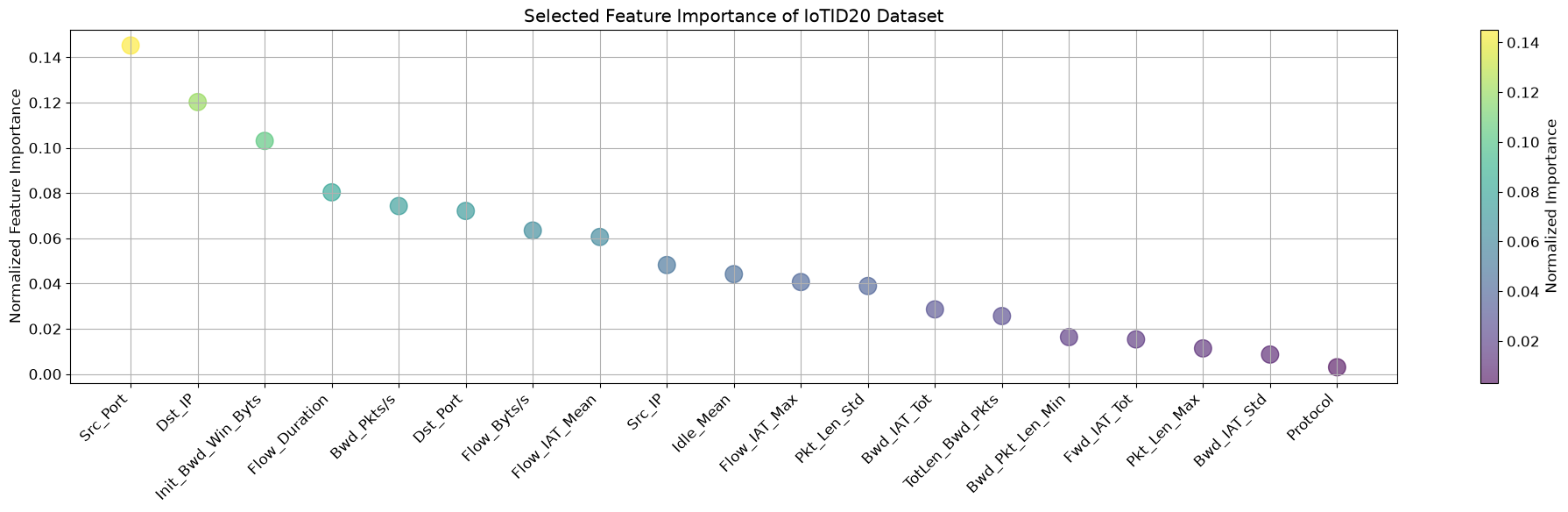}
 \caption{Selected features and normalized importance scores of the final LLM-assisted IDS on IoTID20.} \label{fs_iot}
\end{figure*}

\begin{table*}[tb]
\centering
\renewcommand{\arraystretch}{1.30}
\setlength{\extrarowheight}{1pt}
\setlength{\tabcolsep}{3.6pt}
\caption{Ablation study of the LLM-controlled components under a 10-trial CASH budget.}
\scalebox{0.80}{
\begin{tabular}{|C{1.7cm}|C{4.5cm}|C{1.5cm}|C{1.5cm}|C{1.5cm}|C{1.5cm}|C{1.7cm}|C{2.0cm}|C{1.2cm}|}
\hline
\textbf{Dataset}
& \textbf{Variant}
& \textbf{Test Accuracy (\%)}
& \textbf{Test Precision (\%)}
& \textbf{Test Recall (\%)}
& \textbf{Test F1 (\%)}
& \textbf{Optimization time (s)}
& \textbf{Test time per sample (ms)}
& \textbf{Features} \\ \hline

\multirow{6}{*}{CICIDS2017}
& A0: Traditional AutoML - 10 trials & 99.534 & 99.539 & 99.534 & 99.532 & 37.41 & 0.0145 & 49 \\ \cline{2-9}
& A1: LLM AutoDP only & 99.534 & 99.540 & 99.534 & 99.532 & 33.79 & 0.0137 & 48 \\ \cline{2-9}
& A2: LLM AutoFE only & 99.571 & 99.576 & 99.571 & 99.569 & 34.47 & 0.0178 & 40 \\ \cline{2-9}
& A3: LLM CASH only & 99.608 & 99.612 & 99.608 & 99.606 & 54.07 & 0.0160 & 49 \\ \cline{2-9}
& A4: LLM AutoDP + AutoFE & 99.515 & 99.521 & 99.515 & 99.513 & 33.60 & 0.0136 & 41 \\ \cline{2-9}
& \textbf{A5: Full proposed LLM-AutoML (LLM AutoDP + AutoFE + CASH)} & \textbf{99.683} & \textbf{99.685} & \textbf{99.683} & \textbf{99.680} & 46.55 & 0.0198 & 41 \\ \hline

\multirow{6}{*}{IoTID20}
& A0: Traditional AutoML - 10 trials & 98.834 & 98.869 & 98.834 & 98.844 & 29.66 & 0.0149 & 59 \\ \cline{2-9}
& A1: LLM AutoDP only & 98.834 & 98.869 & 98.834 & 98.844 & 28.69 & 0.0155 & 59 \\ \cline{2-9}
& A2: LLM AutoFE only & 98.866 & 98.897 & 98.866 & 98.875 & 21.99 & 0.0140 & 19 \\ \cline{2-9}
& A3: LLM CASH only & 99.153 & 99.156 & 99.153 & 99.154 & 65.31 & 0.0364 & 59 \\ \cline{2-9}
& A4: LLM AutoDP + AutoFE & 98.866 & 98.897 & 98.866 & 98.875 & 34.18 & 0.0200 & 19 \\ \cline{2-9}
& \textbf{A5: Full proposed LLM-AutoML (LLM AutoDP + AutoFE + CASH)} & \textbf{99.185} & \textbf{99.187} & \textbf{99.185} & \textbf{99.186} & 36.28 & 0.0196 & 19 \\ \hline
\end{tabular}
}
\label{tab:ablation}
\end{table*}

Table~\ref{tab:performance_efficiency} compares the proposed LLM-AutoML framework with optimized deep-learning models~\cite{pso_lstm_ids,ga_cnn_ids}, existing AutoML-based IDSs~\cite{oe_ids,automl_id}, plain tree-based learners~\cite{lightgbm_ids,iot_idsai}, and traditional AutoML baselines~\cite{automl_ccs}. For a fair comparison, all methods were reproduced and evaluated under the same experimental environment, dataset partitions, preprocessing procedure, and evaluation protocol. Traditional AutoML-TPE is evaluated with both 10 and 30 trials: the 10-trial setting provides the direct budget-matched comparison with the proposed method, while the 30-trial setting represents the broader conventional search used as the secondary efficiency reference.

On CICIDS2017, the proposed 10-trial LLM-AutoML method achieves the highest test performance, with 99.683\% accuracy, 99.685\% precision, 99.683\% recall, and 99.680\% weighted F1-score. Under the matched 10-trial budget, Traditional AutoML-TPE reaches 99.532\% F1-score, so the proposed policy improves the test F1-score by 0.149 percentage points. Its optimizer time is 46.55~s compared with 37.41~s for the matched 10-trial TPE run, showing that the equal-budget benefit is primarily better search quality rather than lower raw optimizer time. The broader 30-trial Traditional AutoML-TPE baseline reaches 99.624\% F1-score and requires 128.41~s. Relative to that baseline, the proposed 10-trial method improves F1-score by 0.056 percentage points and reduces optimizer time by 63.7\%. Compared with Traditional AutoML-RS, it improves F1-score by 0.167 percentage points and reduces optimizer time by 58.4\%. An additional 30-trial LLM-assisted run reached the same 99.680\% test F1-score as the 10-trial proposed method while requiring 111.40~s of optimizer time, indicating that the selected compact policy had already converged to the reported hold-out result within the smaller budget.

The same pattern is observed on IoTID20. The proposed method obtains 99.185\% accuracy, 99.187\% precision, 99.185\% recall, and 99.186\% weighted F1-score. Under the equal 10-trial budget, Traditional AutoML-TPE reaches 98.844\% F1-score, giving the proposed method a 0.341-percentage-point advantage, although optimizer time increases from 29.66~s to 36.28~s. Compared with the 30-trial TPE baseline, which reaches 99.170\% F1-score in 72.36~s, the proposed method improves F1-score by 0.016 percentage points and reduces optimizer time by 49.9\%. It also reduces optimizer time by 66.2\% relative to Traditional AutoML-RS while improving F1-score by 0.098 percentage points. As on CICIDS2017, the auxiliary 30-trial LLM-assisted run did not improve the test F1-score beyond 99.186\%, but increased optimizer time to 92.63~s. This result supports the use of the smaller LLM-guided budget for the final method.

The optimized IDS and plain-model comparisons provide additional context. On CICIDS2017, the plain RF baseline is already strong at 99.569\% F1-score, illustrating that the dataset is favorable to tree-based classification. The proposed method nevertheless reaches a higher F1-score while reducing the final feature set from 77 raw features to 41. On IoTID20, the strongest plain RF baseline reaches 99.075\% F1-score, whereas the proposed method reaches 99.186\% with only 19 retained features. The proposed model also remains substantially less expensive to optimize than PSO-LSTM, GA-CNN, OE-IDS, and AutoML-ID in the reported comparison. These results suggest that the main contribution is not a large absolute increase over already strong tree baselines, but a more compact and controlled AutoML process that reaches a strong final model with a small search budget.

The feature-selection results in Figs.~\ref{fs_cic} and~\ref{fs_iot} further show how the final representations differ between the two datasets. CICIDS2017 retains 41 features. Its three largest normalized importance values are associated with \textit{Init\_Win\_bytes\_forward}, \textit{Init\_Win\_bytes\_backward}, and \textit{Flow IAT Min}, which together account for approximately 35.0\% of the total importance of the retained set. The remaining importance is distributed across timing, packet-length, and flag-related features, indicating that the final CICIDS2017 model still uses a relatively broad representation. IoTID20 is reduced more aggressively to 19 features. \textit{Src\_Port}, \textit{Dst\_IP}, and \textit{Init\_Bwd\_Win\_Byts} are the three most important retained features and together account for approximately 36.7\% of the normalized importance. The other retained variables include flow-rate, inter-arrival-time, packet-length, protocol, and acknowledgement-related information. These figures are consistent with the feature counts in Table~\ref{tab:performance_efficiency} and show that the LLM-assisted AutoFE policy does not impose the same reduction level on both datasets.

The inference-time results indicate a modest cost for the final tuned LightGBM configuration. On CICIDS2017, the proposed model requires 0.0198~ms per test sample, which is slower than the plain tree-based baselines and Traditional AutoML-RS, but comparable to the 30-trial Traditional AutoML-TPE model at 0.0208~ms. On IoTID20, the proposed model requires 0.0196~ms per sample, which is higher than the plain tree-based and Traditional AutoML baselines but remains below 0.1~ms per sample. Thus, the proposed method trades a small amount of per-sample inference time for higher test F1-score and a substantially smaller final feature set, particularly on IoTID20.

Table~\ref{tab:ablation} expands the ablation study to the same performance and efficiency measures used in the main comparison. All variants use a 10-trial CASH budget, so the table isolates how the LLM-controlled AutoDP, AutoFE, and CASH components interact under a fixed search budget. On CICIDS2017, A1 (LLM AutoDP only) is effectively unchanged from A0 in F1-score, while A2 (LLM AutoFE only) increases F1-score from 99.532\% to 99.569\% and reduces the feature count from 49 to 40. A3 (LLM CASH only) provides the largest individual performance increase, reaching 99.606\% F1-score. A4, which combines LLM AutoDP and AutoFE but retains the traditional CASH policy, drops slightly to 99.513\% F1-score. The full A5 method then reaches 99.680\%, clearly above every partial variant. This non-additive pattern indicates that the transformed data and feature space benefit from a CASH policy that is aligned with them; the gains from the three controls should therefore be interpreted as coordinated rather than independent additive effects.

IoTID20 shows an even clearer separation of roles. A1 produces the same 98.844\% F1-score as A0, indicating that the LLM AutoDP policy alone does not change the final detection result in this run. A2 improves F1-score slightly to 98.875\% while reducing the feature set from 59 to 19. A3 provides the dominant individual performance gain and reaches 99.154\% F1-score, although its optimizer time is the highest among the partial variants at 65.31~s. A4 retains the compact 19-feature representation but does not improve beyond A2. When all three components are coordinated in A5, F1-score rises to 99.186\%, the highest result in the ablation table. The ablation therefore suggests that CASH guidance contributes most directly to predictive performance, whereas AutoFE contributes strongly to compactness and becomes most effective when coupled with the final LLM-guided search.

Taken together, the equal-budget and ablation results refine the efficiency claim of this work. The LLM-assisted policy is not faster than Traditional AutoML-TPE when both are restricted to exactly 10 trials; it incurs modest additional optimizer cost while reaching a clearly stronger test result on both datasets. Its efficiency advantage appears when the proposed compact 10-trial search is compared with the broader 30-trial conventional search, where it reduces optimizer time by approximately 63.7\% on CICIDS2017 and 49.9\% on IoTID20 while maintaining slightly higher F1-score. The results therefore support the bounded-policy architecture as a way to improve low-budget search quality and reduce the amount of broader AutoML exploration required, while final model selection remains governed by cross-validation and independent hold-out testing.

\section{Conclusion} \label{S5}
AutoML has emerged as a promising approach to autonomous cybersecurity for IoT and zero-touch networks, but the cost of broad model and hyperparameter search remains an important limitation. This paper proposed an LLM-assisted AutoML framework for IoT intrusion detection in which a local Llama~3.1 8B model generates bounded and validated policies for data balancing, feature selection, and CASH optimization. The LLM does not train or evaluate the IDS model; its role is limited to policy generation, while the resulting configurations are executed and assessed through deterministic ML procedures, cross-validation, and an independent hold-out test set.
The final experiments clarify both the effectiveness and the efficiency of this design. Under the same 10-trial budget, the proposed policy increases weighted test F1-score from 99.532\% to 99.680\% on CICIDS2017 and from 98.844\% to 99.186\% on IoTID20 relative to Traditional AutoML-TPE. The equal-budget LLM-assisted search requires modestly more optimizer time, but it reaches a stronger low-budget result. Relative to the broader 30-trial Traditional AutoML-TPE baseline, the proposed 10-trial method reduces optimizer time by 63.7\% on CICIDS2017 and 49.9\% on IoTID20 while maintaining slightly higher F1-score. The ablation study further shows that the three LLM-controlled stages interact non-additively: CASH guidance contributes most directly to detection performance, whereas AutoFE substantially reduces the retained feature set, and their coordinated use produces the strongest final result. These findings support the bounded-policy architecture as a practical way to improve the quality of a compact AutoML search without placing the LLM in the deployed IDS decision loop.
Future work will extend the framework to online and continual IDS learning, evaluate additional local LLMs, and incorporate multi-objective deployment constraints such as latency, memory, and energy consumption.

\section{Acknowledgment}
This work was supported in part by the Natural Sciences and Engineering Research Council of Canada (NSERC) under Discovery Grant RGPIN-2025-05840, and in part by the Department of National Defence (DND)/NSERC Discovery Grant Supplement DGDND-2025-05840.

\begin{IEEEbiography}[{\includegraphics[width=1in,height=1.25in,clip,keepaspectratio]{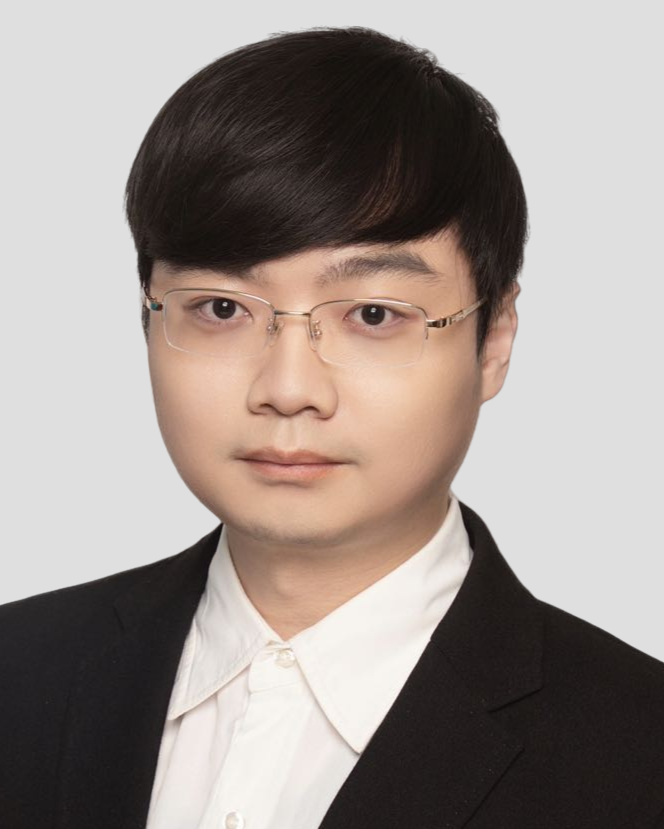}}]{Li Yang}\, (Member, IEEE) is an Assistant Professor in the Faculty of Business and Information Technology at Ontario Tech University, and an Adjunct Research Professor in the Department of Electrical and Computer Engineering at Western University. He received his Ph.D. in Electrical and Computer Engineering from Western University in 2022. He was awarded the Faculty of Business and Information Technology (FBIT) Early Career Research Chair in Cybersecurity for Business, the FBIT Research Excellence Award, and the Ontario Tech University Research Excellence Award in 2026. Li Yang has been named by the IEEE Computer Society as one of the recipients of Computing's Top 30 Early Career Professionals for 2025. Li Yang is also included in Stanford University/Elsevier's List of the World's Top 2\% Scientists, and he was ranked among the world's top 0.5\% of researchers in 'Networking \& Telecommunications' in 2024 and 2025. He is currently an Associate Editor of IEEE Transactions on Industrial Informatics and IEEE Transactions on Network and Service Management. His research interests include cybersecurity, machine learning, deep learning, AutoML, Tiny Machine Learning (TinyML), LLMs, model optimization, network data analytics, IoT, intrusion detection, anomaly detection, concept drift, continual learning, and adversarial machine learning.
\end{IEEEbiography}

\end{document}